\documentclass[twocolumn]{aastex631}

\usepackage{multirow}
\usepackage{comment}

\accepted{in ApJL - March 2025}
\shorttitle{Spectrum of a cold exoplanet around a white dwarf.}
\shortauthors{Voyer et al. 2025}

\begin{document}

\title{MIRI-LRS spectrum of a cold exoplanet around a white dwarf: water, ammonia, and methane measurements.}

% Main contributors
\author[0000-0002-0615-9253]{Maël Voyer}
\affiliation{Université Paris Cité, Université Paris-Saclay, CEA, CNRS, AIM, F-91191 Gif-sur-Yvette, France}
\author[0000-0001-6516-4493]{Quentin Changeat}
\affiliation{Kapteyn Institute, University of Groningen, 9747 AD Groningen, NL.}
\affiliation{Department of Physics and Astronomy, University College London, Gower Street, WC1E 6BT, UK}
\author{Pierre-Olivier Lagage}
\affiliation{Université Paris-Saclay, Université Paris Cité, CEA, CNRS, AIM, F-91191 Gif-sur-Yvette, France}

\author{Pascal Tremblin}
\affiliation{Université Paris-Saclay, UVSQ, CNRS, CEA, Maison de la Simulation, 91191, Gif-sur-Yvette, France}

% co-PIS
\author[0000-0002-5462-9387]{Rens Waters}
\affiliation{SRON Netherlands Institute for Space Research, Niels Bohrweg 4, 2333 CA Leiden, the Netherlands} %2
\affiliation{Department of Astrophysics/IMAPP, Radboud University, PO Box 9010, 6500 GL Nijmegen, the Netherlands} %11
\affiliation{HFML - FELIX. Radboud University, PO Box 9010, 6500 GL Nijmegen, the Netherlands} %12
\author[0000-0001-9818-0588]{Manuel G\"udel}
\affiliation{Department of Astrophysics, University of Vienna, Türkenschanzstrasse 17, 1180 Vienna, Austria} %8
\affiliation{ETH Zürich, Institute for Particle Physics and Astrophysics, Wolfgang-Pauli-Strasse 27, 8093 Zürich, Switzerland} %9
\author[0000-0002-1493-300X]{Thomas Henning}
\affiliation{Max-Planck-Institut für Astronomie (MPIA), Königstuhl 17, 69117 Heidelberg, Germany} %4

% co-Is exoplanet group  + full members
\author[0000-0002-4006-6237]{Olivier Absil}
\affiliation{STAR Institute, Université de Liège, Allée du Six Août 19c, 4000 Liège, Belgium} %16
\author[0000-0002-5971-9242]{David Barrado}
\affiliation{Centro de Astrobiología (CAB), CSIC-INTA, ESAC Campus, Camino Bajo del Castillo s/n, 28692 Villanueva de la Cañada, Madrid, Spain} %17
\author[0000-0001-9353-2724]{Anthony Boccaletti}
\affiliation{LESIA, Observatoire de Paris, Université PSL, CNRS, Sorbonne Université, Univ. Paris Diderot, Sorbonne Paris Cité, 5 place Jules Janssen, 92195 Meudon, France} %18
\author[0000-0003-4757-2500]{Jeroen Bouwman}
\affiliation{Max-Planck-Institut für Astronomie (MPIA), Königstuhl 17, 69117 Heidelberg, Germany} %4
\author[0000-0001-6492-7719]{Alain Coulais}
\affiliation{LERMA, Observatoire de Paris, Université PSL, Sorbonne Université, CNRS, Paris, France} %20
\affiliation{Université Paris-Saclay, Université Paris Cité, CEA, CNRS, AIM, F-91191 Gif-sur-Yvette, France} %6
\author[0000-0002-5342-8612]{Leen Decin}
\affiliation{Institute of Astronomy, KU Leuven, Celestijnenlaan 200D, 3001 Leuven, Belgium} %3
\author[0000-0001-9250-1547]{Adrian M. Glauser}
\affiliation{ETH Zürich, Institute for Particle Physics and Astrophysics, Wolfgang-Pauli-Strasse 27, 8093 Zürich, Switzerland} %9
\author[0000-0002-0932-4330]{John Pye}
\affiliation{School of Physics \& Astronomy, Space Park Leicester, University of Leicester, 92 Corporation Road, Leicester, LE4 5SP, UK} %10
\author[0000-0002-2041-2462]{Alistair Glasse}
\affiliation{UK Astronomy Technology Centre, Royal Observatory, Blackford Hill, Edinburgh EH9 3HJ, UK} %21
\author[0009-0007-5200-1362]{René Gastaud}
\affiliation{Université Paris-Saclay, CEA, Département d'Electronique des Détecteurs et d'Informatique pour la Physique, 91191, Gif-sur-Yvette, France} %19
\author[0000-0002-7612-0469]{Sarah Kendrew}
\affiliation{European Space Agency, Space Telescope Science Institute, Baltimore, Maryland, USA} %23
\author[0000-0001-8718-3732]{Polychronis Patapis}
\affiliation{ETH Zürich, Institute for Particle Physics and Astrophysics, Wolfgang-Pauli-Strasse 27, 8093 Zürich, Switzerland} %9
\author[0000-0002-2352-1736]{Daniel Rouan}
\affiliation{LESIA, Observatoire de Paris, Université PSL, CNRS, Sorbonne Université, Univ. Paris Diderot, Sorbonne Paris Cité, 5 place Jules Janssen, 92195 Meudon, France} %18
\author[0000-0001-7591-1907]{Ewine F. van Dishoeck}
\affiliation{Leiden Observatory, Leiden University, P.O. Box 9513, 2300 RA Leiden, the Netherlands} %5
\author[0000-0002-3005-1349]{G\"oran \"Ostlin}
\affiliation{Department of Astronomy, Oskar Klein Centre, Stockholm University, 106 91 Stockholm, Sweden} %27
\author[0000-0002-2110-1068]{Tom P. Ray}
\affiliation{Astronomy \& Astrophysics Section, School of Cosmic Physics, Dublin Institute for Advanced Studies, 31 Fitzwilliam Place, Dublin, D02 XF86, Ireland} %28
\author[0000-0001-7416-7936]{Gillian Wright}
\affiliation{UK Astronomy Technology Centre, Royal Observatory Edinburgh, Blackford Hill, Edinburgh EH9 3HJ, UK} %29

\begin{abstract}
The study of the atmosphere of exoplanets orbiting white dwarfs is a largely unexplored field. With WD\,0806-661\,b, we present the first deep dive into the atmospheric physics and chemistry of a cold exoplanet around a white dwarf. We observed WD 0806-661 b using JWST's Mid-InfraRed Instrument Low-Resolution Spectrometer (MIRI-LRS), covering the wavelength range from 5 -- 12~$\mu \rm{m}$, and the Imager, providing us with 12.8, 15, 18 and 21\,$\mu$m photometric measurements. We carried the data reduction of those datasets, tackling second-order effects to ensure a reliable retrieval analysis. Using the \textsc{TauREx} retrieval code, we inferred the pressure-temperature structure, atmospheric chemistry, mass, and radius of the planet. The spectrum of WD 0806-661 b is shaped by molecular absorption of water, ammonia, and methane, consistent with a cold Jupiter atmosphere, allowing us to retrieve their abundances. From the mixing ratio of water, ammonia and methane we derive $\rm{C/O} = 0.34 \pm 0.06$, $\rm{C/N} = 14.4 ^{+2.5}_{-1.8}$ and $\rm{N/O} = 0.023 \pm 0.004$ and the ratio of detected metals as proxy for metallicity. We also derive upper limits for the abundance of CO and $\rm{CO_2}$ ($1.2\cdot10^{-6} \rm{\,and\,} 1.6\cdot10^{-7}$ respectively), which were not detected by our retrieval models. While our interpretation of WD\,0806-661\,b's atmosphere is mostly consistent with our theoretical understanding, some results -- such as the lack of evidence for water clouds, an apparent increase in the mixing ratio of ammonia at low pressure, or the retrieved mass at odds with the supposed age -- remain surprising and require follow-up observational and theoretical studies to be confirmed.

\end{abstract}

\section{Introduction} \label{sec:intro}

Due to its unprecedented sensitivity and unique wavelength coverage, JWST has opened a new window into the characterization of Y-dwarfs ($\rm{T_{eff}} < 500~\rm{K}$) and low-temperature exoplanets \citep{2023PASP..135d8001R, BEILER_SED_YDWARF_NIR_2_MIR, Barrado_15NH3, 2024ApJ...971..121K, 2024arXiv241010933K}. Previously, such objects were mainly observed using ground-based instruments, and mid-infrared spectroscopy was unavailable. 
Recently, studies of VHS-1256-1257\,b by JWST demonstrated the pivotal capabilities of the telescope in the infrared, and its accurate absolute flux calibration \citep{Miles_BD_ground_2020, Miles_ERS_VHS}. 
JWST enables spectroscopic observations of much colder and fainter planetary-mass self-luminous objects -- of similar characteristics to solar-system objects -- which is of particular interest to the exoplanet community.

One such intriguing object is WD\,0806-661\,b (hereafter WD\,0806\,b), a planetary-mass companion orbiting a white dwarf (WD) at an apparent separation of 2500 au \citep{Luhman_wd-0806_discov, Luhman_wd-0806_confirm}. 
Planetary systems around WDs have been the subject of many studies \citep{2021orel.bookE...1V, 2024arXiv240307427M}. Polluted WDs, debris disks, and dust disks allow us to study the composition of destroyed exoplanets and planetesimals \citep{ 2012MNRAS.424..333G, 2019AJ....158..242X, 2024MNRAS.532.3866R}. Yet, atmospheres of planets around WDs remain difficult to characterize, with so far only a single directly observed exo-atmosphere in a WD system (WD\,1856+534\,b). This particular object did not show molecular features \citep{2021A&A...649A.131A, 2021AJ....162..296X}. Accordingly, WD\,0806\,b is a unique opportunity the atmospheres of exoplanets around WDs.

Based on evolutionary models \citep{Burrows_th_BD_model, Saumon_evo_L&T_Dwarf} using the system age of $2\pm0.5$\,Gyr \citep{Subasavage_wd0806_age}, the comparison of observed magnitudes in J-band ($\rm{m_{J}}$) and in a Spitzer-band ($\rm{m_{ [4.5]}}$) suggests that WD\,0806\,b has an effective temperature of $\rm{T_{eff}} \sim 300 - 345$\,K and a mass of $\rm{M_{p}} \in [6, 9]\,\rm{M_J}$ \citep{Luhman_wd-0806_confirm}. In this letter, we present a newly acquired spectrum of WD\,0806\,b, obtained by the James Webb's MIRI Low-Resolution Spectrometer (LRS) and the F1280W, F1500W, F1800W and F2100W filters of MIRI's Imager \citep{Rieke_MIRI_Intro,2023PASP..135d8003W,Bouchet_MIRIM}.  The data were acquired as part of the GTO 01276, (PI: Lagage) on the 14$^{\rm{th}}$ of July 2023. WD\,0806\,b is the first planetary-mass object around a white dwarf to unveil molecular features through direct spectroscopy.

Our analysis of WD\,0806\,b starts from the raw data. Our data reduction pipeline, which relies on the official STScI steps along with customized functions, and our retrieval setup are described in Section \ref{sec:methodo}. Our retrieval results are shown in Section \ref{sec:Results}. Finally, Section \ref{sec:Discussion} provides additional discussion on the implications and robustness of our findings.\\

\begin{figure*}[ht!]
\includegraphics[width=1\textwidth]{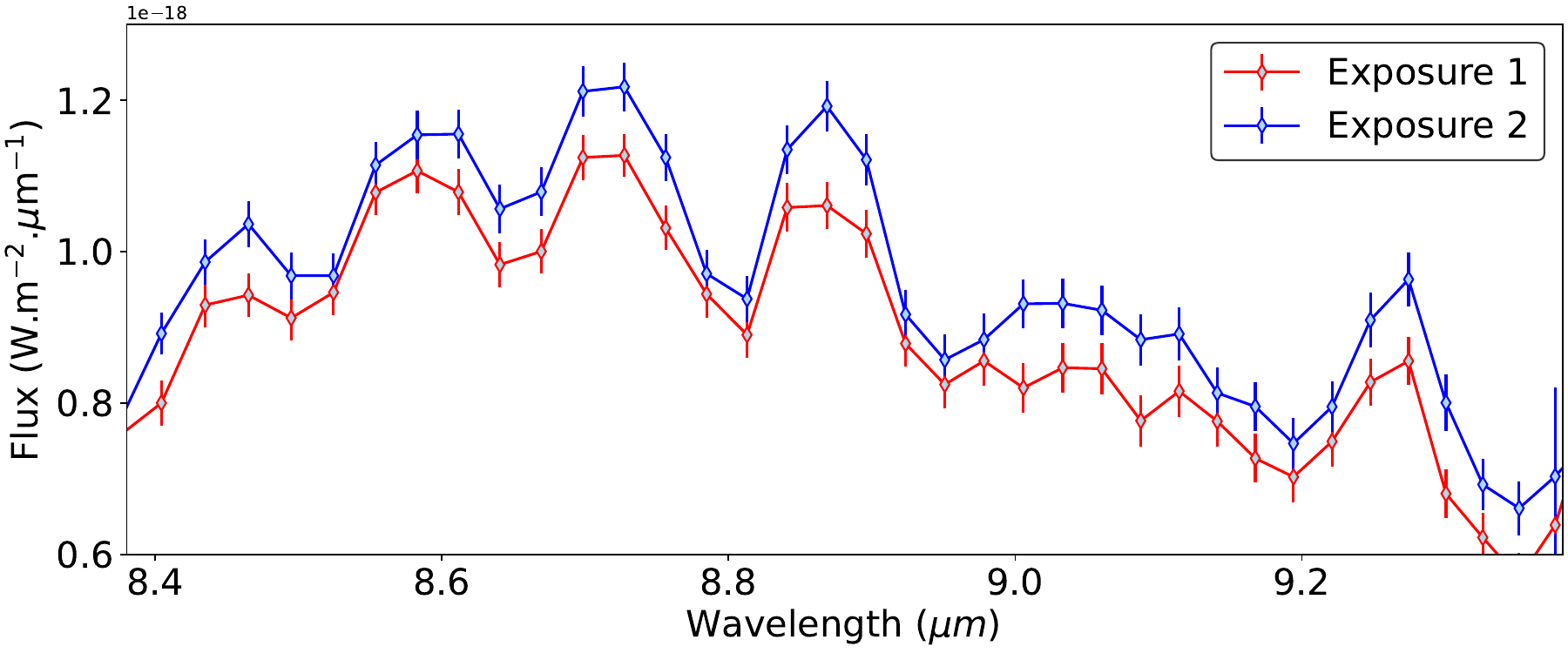}
\caption{Extracted spectra from exposure 1 (red) and exposure 2 (blue). We show a zoom of the spectra for better visibility.
\label{fig:flux_offset}}
\end{figure*}

\section{Methodology} \label{sec:methodo}

\subsection{Observations}\label{sec:methodo:1}

The observations of WD\,0806\,b were taken by the MIRI-LRS (R $\approx100$) instrument using a two-dither pattern, the source being too faint for MIRI-MRS. Each slit position, was exposed for 1.79 hours (one exposure with eight integrations of 290 groups in FASTR1 mode). The availability of two slit positions is in principle advantageous for background and detector systematics subtraction. The standard practice is to subtract the detector images from the two exposures, leaving a positive and negative spectral trace.
Using MIRI Imager, we also observed WD\,0806\,b with the 12.8, 15, 18 and 21 microns filters for 277, 560, 283 and 283 seconds respectively (i.e., $50$, $50$, $25$ and $25$ groups in $1$, $2$, $2$, and $2$ integrations using the FASTR1 mode).

\subsection{Data reduction of the MIRI-LRS data}

For Stages 1 and 2 of the data reduction, we mostly utilize the official JWST STScI pipeline. We used the most recent version at the time of the reduction: pipeline 1.12.5 and the CRDS (Calibration
Reference Data System) 11.17.14. However, we modified the steps pertaining to error propagation, outlier rejection, background subtraction, and wavelength calibration to optimally process our data, and we did not use Stage 3 at all. \\

- {\it Stage 1}: The first stage (i.e., detector level correction) is setup using the default parameters of the STScI pipeline, except for the reset step which is merged with the dark current subtraction step. We then use the Stage 1 output rateints files to perform a customized background subtraction. Details are described in Appendix \ref{Appendix:A1}. \\

- {\it Stage 2}: After the custom background subtraction, we apply the default Stage 2 step from the STScI JWST pipeline, providing us with the calints files. We perform our outlier detection and interpolation at the end of Stage $2$. Details are in Appendix \ref{Appendix:A2}  \\

- {\it New step, “background rescaling”}: We add a new step at the end of the stage 2 to rescale inconsistent flux levels (i.e., offsets) between the two slit positions. Figure \ref{fig:flux_offset} shows the flux offsets originating from the background before the application of this step.
Integrations within the same exposures are consistent, this suggests that the discrepancy arises from a varying background between the two slit positions.
Discussions with instrument experts at STScI have led to the hypothesis that light coming from the MIRI Imager -- which is close to saturation at the end of each integration -- could spread to the LRS subarray differently due to the slight change in field of view (i.e., $\sim 1.5^"$)\footnote{Our dataset is dominated by a background with a right to left gradient (see Fig. \ref{fig:appendixA2}). However, zodiacal light -- the dominant source of sky background for $\lambda \in [5,12]\,\mu$m \citep{2003Icar..164..384R, 2010A&A...523A..53P} -- is expected to be spatially constant in the 2.4 arcsec$^2$ field of view of the LRS slit \citep{2012ApJ...760..102P}, and therefore cannot explain the changing background.}

In preparation for further statistical analysis (i.e., atmospheric retrievals), this issue needs to be corrected: for direct imaging, absolute flux and non-astrophysical wavelength dependent signals can strongly bias parameter estimates. To remove the background, we make use of the two dither positions (i.e., as done by the default STScI step). As shown in Figure \ref{fig:flux_offset}, too much background appears to be subtracted for the first dither, and too little for the second dither. We perform a rescaling correction by applying a wavelength dependent factor inferred by fitting a second order polynomial to the median background pixels (see Appendix \ref{Appendix:A3}). \\

\begin{figure*}[ht!]
\includegraphics[width=1\textwidth]{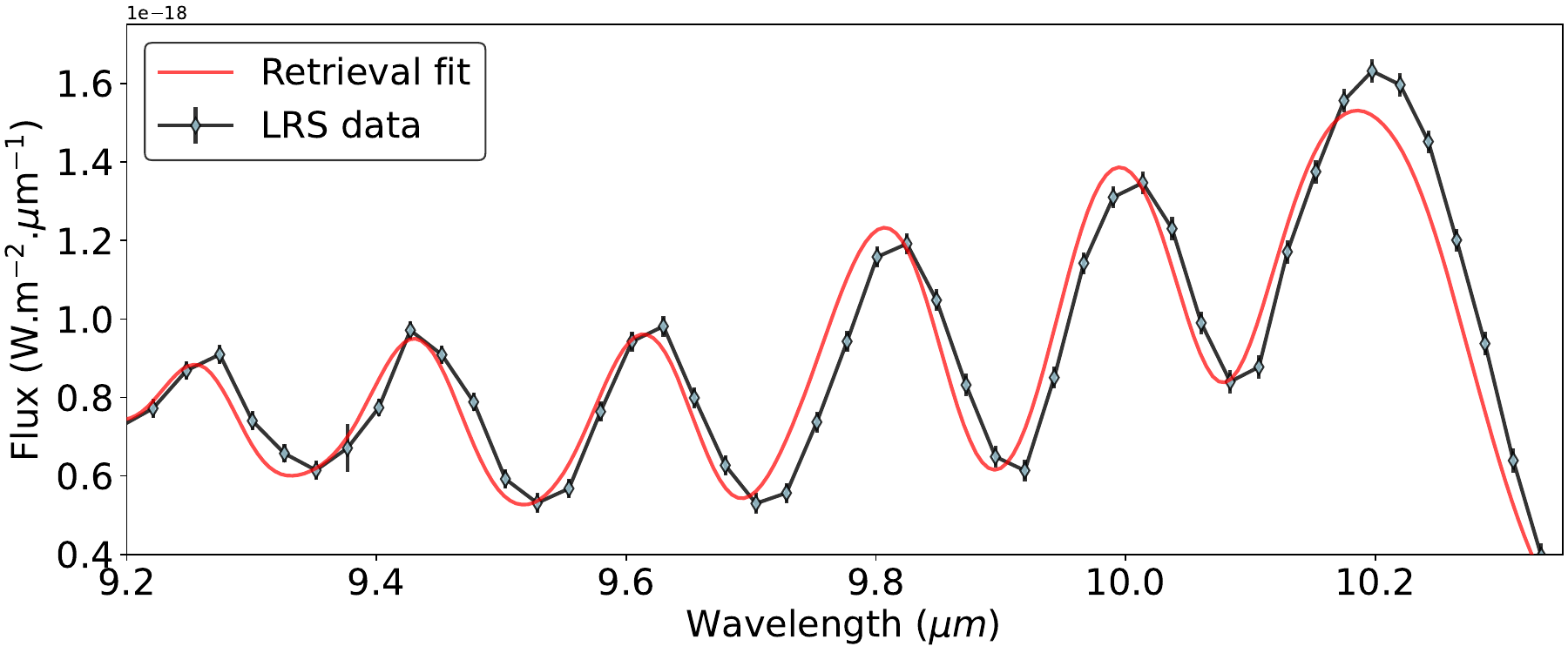}
\caption{LRS spectrum using the default wavelength calibration (black) and best fit from TauRex (red). A wavelength discrepancy ($\Delta\lambda \sim 10$\,nm)  is clearly visible.
\label{fig:wl_offset}}
\end{figure*}

- {\it Stage 3}: We apply a customized Stage 3 step to extract and combine the spectra from the different images. Our extraction aperture is a rectangular, six pixels wide, box centered on the spectral trace. The step is performed  {\it individually} for each integration and several corrections are applied (the details are available in Appendix \ref{Appendix:A4}).

\subsection{Data reduction of the MIRI filters data}\label{sec:methodo:2}

Additionally, four observations of WD\,0806\,b were obtained with the MIRI filters F1280, F1500, F1800 and F2100. We use the STScI pipeline to reduce these datasets and get the “cal” files. The source at 21 $\mu$m could not be detected by the STScI pipeline. We did detect it using a custom flux extraction, background subtraction and aperture correction. Using Photutils we used a circular aperture to sum the flux coming from the source and an annulus aperture to measure the background level. The radii of these apertures were taken from the JWST CRDS file $jwst\_miri\_apcorr\_0010$ for proper aperture correction. For consistency, this custom step is applied to all four filters. In the cases where the STScI pipeline can detect the target, our extracted fluxes using the custom step are within the uncertainties of the values from the STScI pipeline. For  F1280, F1500, F1800 and F2100 we obtain the photometric fluxes $F_\mathrm{p} = \{5.8 \pm 0.1 
 ,\, 4.75 \pm 0.08 ,\, 3.0 \pm 0.1,\, 2.1 \pm 0.3 \} \times 10^{-19}\, \rm{W.m^{-2}.\mu m^{-1}}  $ respectively.

\subsection{Atmospheric retrieval}\label{sec:methodo:3}

Once the spectrum is obtained, we extract the information content using the \textsc{TauREx3} atmospheric retrieval code \citep{Al-Refaie_taurex3_2021, Al-Refaie_taurex3_2022}. 

The retrievals are performed on the {\it LRS data only} since incompatibilities between datasets -- from planet variability \citep{Miles_ERS_VHS, 2024ApJS..270...34C} and instrument systematics \citep{2021AJ....161....4Y, 2024RASTI...3..415E}  -- are known to be possible. At native pixel resolution, spectral lines are spread by the optics onto several pixels in the spectral direction. Analyzing  the data at native resolution, we need to account for that effect. We utilize the TauREx-multimodel\footnote{The code and documentation can be found at : \url{https://github.com/groningen-exoatmospheres/taurex-multimodel}} plugin \citep{Changeat_2025}, which allows convolving the high-resolution theoretical spectrum with the line spread function of the instrument.

\begin{figure*}[ht!]
\includegraphics[width=1\textwidth]{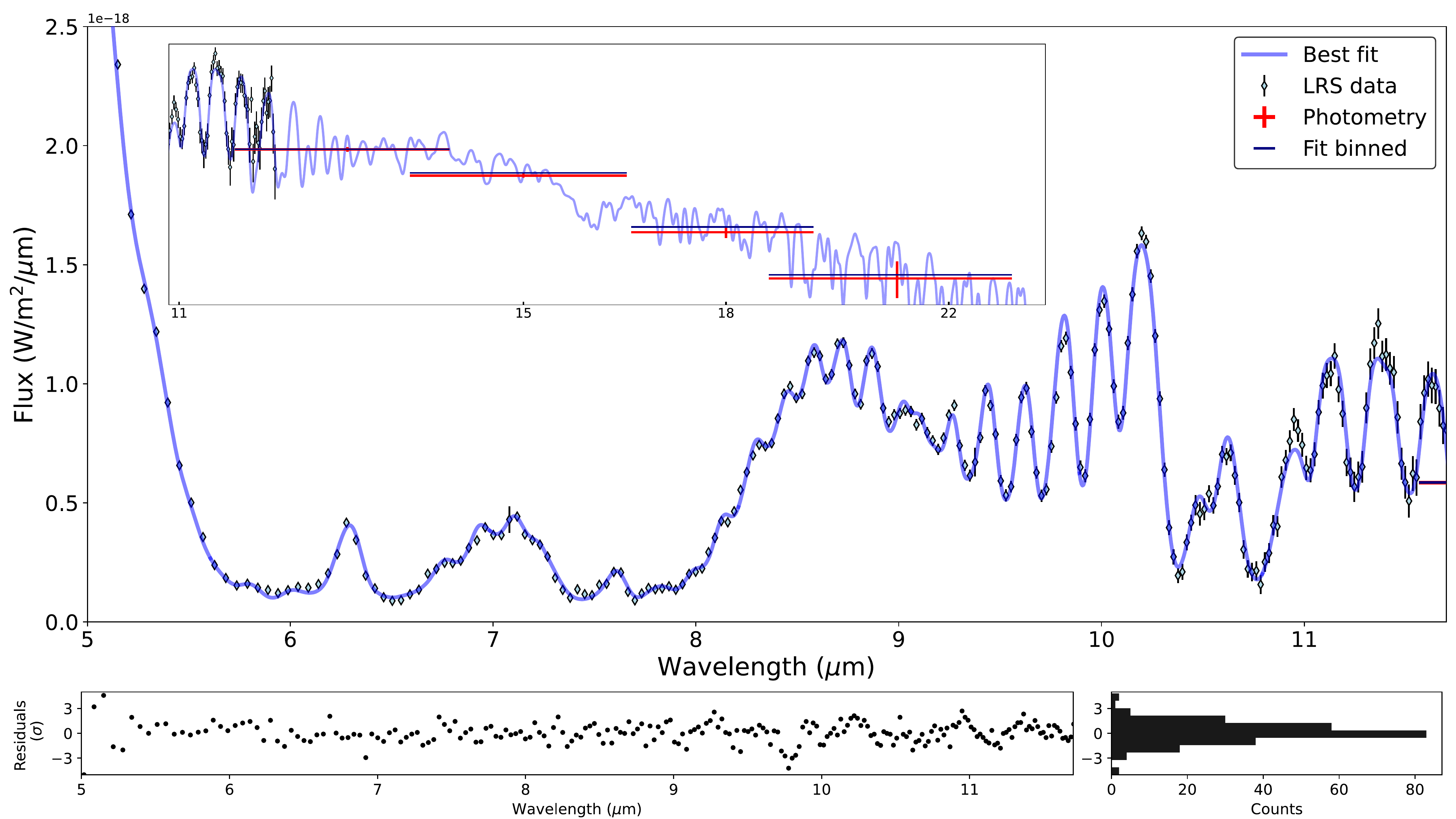}
\caption{Top panel : MIRI fixed slit LRS spectrum of WD\,661\,b (black) and best fit model from TauREx (blue). The black rectangle in the top left-hand corner shows the $\lambda = [11, 24]\,\rm{\mu m}$ range. The red lines are the $\lambda \in \{12.8, 15, 18, 21\}\,\mu$m photometric points from MIRI. They are not included in the retrieval, just plotted over the best fit. The navy line at the same places corresponds to the best fit, binned to the resolution of these photometric points. The bottom left-hand panel displays the residuals between the LRS data and the best fit, for which the root-mean-standard-deviation is $1.247$}. The bottom right-hand panel shows the histograms of these residuals.
\label{fig:LRS&fit}
\end{figure*}

The planet is modeled with a plane-parallel atmosphere of 80 layers spanning the pressures $\mathrm{p} \in [10^6,0.1]$\,Pa in log space, assuming hydrostatic equilibrium. We use a hydrogen/helium atmosphere with a solar ratio of 0.17. For radiatively active species, we use the cross-sections at resolution $R=50\,000$ from the ExoMol project \citep{Tennyson_ExoMol_2016, Chubb_ExoMol_2021}. We included absorption from water \citep{polyansky_exomol_2018}, ammonia \citep{2011MNRAS.413.1828Y}, methane \citep{2014MNRAS.440.1649Y}, carbon monoxide \citep{2015ApJS..216...15L}, carbon dioxide \citep{2020MNRAS.496.5282Y}, and phosphine \citep{2015MNRAS.446.2337S}. Each abundance is freely retrieved with molecular mixing ratio (MR) either constant-with-altitude, or with a two-layer profile \cite[see:][]{2019ApJ...886...39C}. We include Collision Induced Absorption (CIA) from H2-H2 \citep{Abel_CIA_H2H2_2011, Fletcher_CIA_H2H2_2018}  and H2-He \citep{Abel_CIA_H2He_2012},  and Rayleigh scattering \citep{Cox_rayleight_2015}. Aerosols are considered using the TauREx-PyMieScatt\footnote{The code and documentation can be found at : \url{https://github.com/groningen-exoatmospheres/taurex-pymiescatt}} plugin \citep{Changeat_2025}. TauREx-PyMieScatt allows the inclusion of parameterized aerosols using their wavelength dependent optical constants. Given the predicted and retrieved thermal structure (see Fig. \ref{fig:LRS&fit}) we considered water ice clouds, and ammonia ice clouds \citep{article,2007JOSAB..24..126H} -- similar in composition to Jupiter's clouds -- and fully opaque grey clouds. In this study, condensate species such as salt clouds and silicate clouds, commonly often discussed in brown-dwarfs studies, were excluded since our temperature regimes are much colder.
For the parameter exploration, we used uniform priors as described in Appendix \ref{Appendix : B} and sampled the parameter space using the Nested Sampling algorithm MultiNest \citep{Feroz_MultiNest_2009, Buchner_MultiNest_2014}. MultiNest is set up with an evidence tolerance of 0.5 and 2000 live points. We here define the core retrievals for our analysis: \\

- {\it Baseline setup}:  $\rm{H_2O}$, $\rm{NH_3}$ and $\rm{CH_4}$ constant MR through the atmosphere.\\

- {\it Best setup}:  $\rm{H_2O}$ and $\rm{CH_4}$ constant through the atmosphere. Two-layer $\rm{NH_3}$. \\

- {\it 5 species setup} : $\rm{H_2O}$, $\rm{CH_4}$, $\rm{CO}$ and $\rm{CO_2}$ constant through the atmosphere. Two-layer $\rm{NH_3}$. No other molecule. \\

- {\it Cloudy setup}:  $\rm{H_2O}$ and $\rm{CH_4}$ constant through the atmosphere. Two-layer $\rm{NH_3}$. $\rm{H_2O}$ ice and $\rm{NH_3}$ ice clouds. \\

- {\it Free water setup}: Constant $\rm{CH_4}$. Two-layer $\rm{NH_3}$ and  $\rm{H_2O}$ \footnote{The atmosphere is too hot for methane condensation, so we did not explore two-layer profile for that specie}.

\subsection{Wavelength Calibration}\label{sec:methodo:4}

In our first set of retrievals, results showed a definite, wavelength-dependent mismatch between line positions of the data and the theoretical models (see Fig. \ref{fig:wl_offset}, bottom panel). This phase shift is below the current wavelength calibration level of MIRI-LRS (i.e., $\Delta\lambda \sim 20$\,nm), and therefore consistent with expected instrument design specifications. Contributions to this mismatch between data and theoretical spectra could also come from uncertainties in the cross-sections, albeit unlikely to be of major importance. In any case, the spectra need a wavelength correction to ensure that the inferred atmospheric parameters are not biased by this effect. We therefore fitted for the wavelength solution using a third\footnote{We tested polynomials with orders from one to five, and found that third order is the lowest order that could best explain the data.} order polynomial correction. The best-fit correction for our data is given by: 
\begin{equation}
    \lambda_{shifted} = -1.15\cdot10^{-5} \cdot \lambda^3 + \lambda,
\end{equation}
where $\lambda_{shifted}$ is the corrected wavelength solution and $\lambda$ the unmodified one, both in microns. We provide a table with the adopted wavelength solution in appendix \ref{Appendix : wavelength correction}, Table \ref{tab:wavelength_correction}.

We then fixed that correction for all retrievals for consistency. This correction results in much higher Bayesian evidence (hereafter $E$) with $\Delta ln(E) \approx 300$ compared to the same fits without correction, and impacts the retrieved parameters. Given the Bayesian evidence difference, we emphasize on the need for precise calibration and systematic checks when performing atmospheric retrieval exercises with JWST data.

\section{Results} \label{sec:Results}

The spectrum shape of WD\,0806\,b is as expected and consistent with other observed Y$0$ dwarfs \citep{2019ApJ...877...24Z, BEILER_SED_YDWARF_NIR_2_MIR, 2024ApJ...973..107B, Barrado_15NH3, 2024arXiv241010933K}. We have clear detection of three molecules, NH$_3$, H$_2$O and CH$_4$ with $\Delta \rm{ln(E)} = 5504$, $\Delta \rm{ln(E)} = 3540$, $\Delta \rm{ln(E)} = 15364$ respectively (see Fig. \ref{fig:mol_depth}, \cite{bayes_factor}).

Figure \ref{fig:LRS&fit} shows the results from the “best setup” described in section \ref{sec:methodo:4}, which we utilize in this section. Note that this retrieval uses the MIRI-LRS data from $\lambda \in [5, 11.7]\, \mu$m only. The photometric data ($\lambda \in \{12.8, 15, 18, 21\}\,\mu\rm{m}$) is also plotted in the inset of figure \ref{fig:LRS&fit} to showcase the predictive power of the fit at longer wavelengths. The photometric data matches the forward model at, $0.12,~1.4,~0.95~\rm{and}~0.20~\sigma$ respectively, providing us with additional confidence in our retrieval results. 

In this section, we present the consistent results (within  $3\sigma$), across all setups outlined in Section \ref{sec:methodo:4}.

\begin{figure*}[ht!]
\includegraphics[width=1\textwidth]{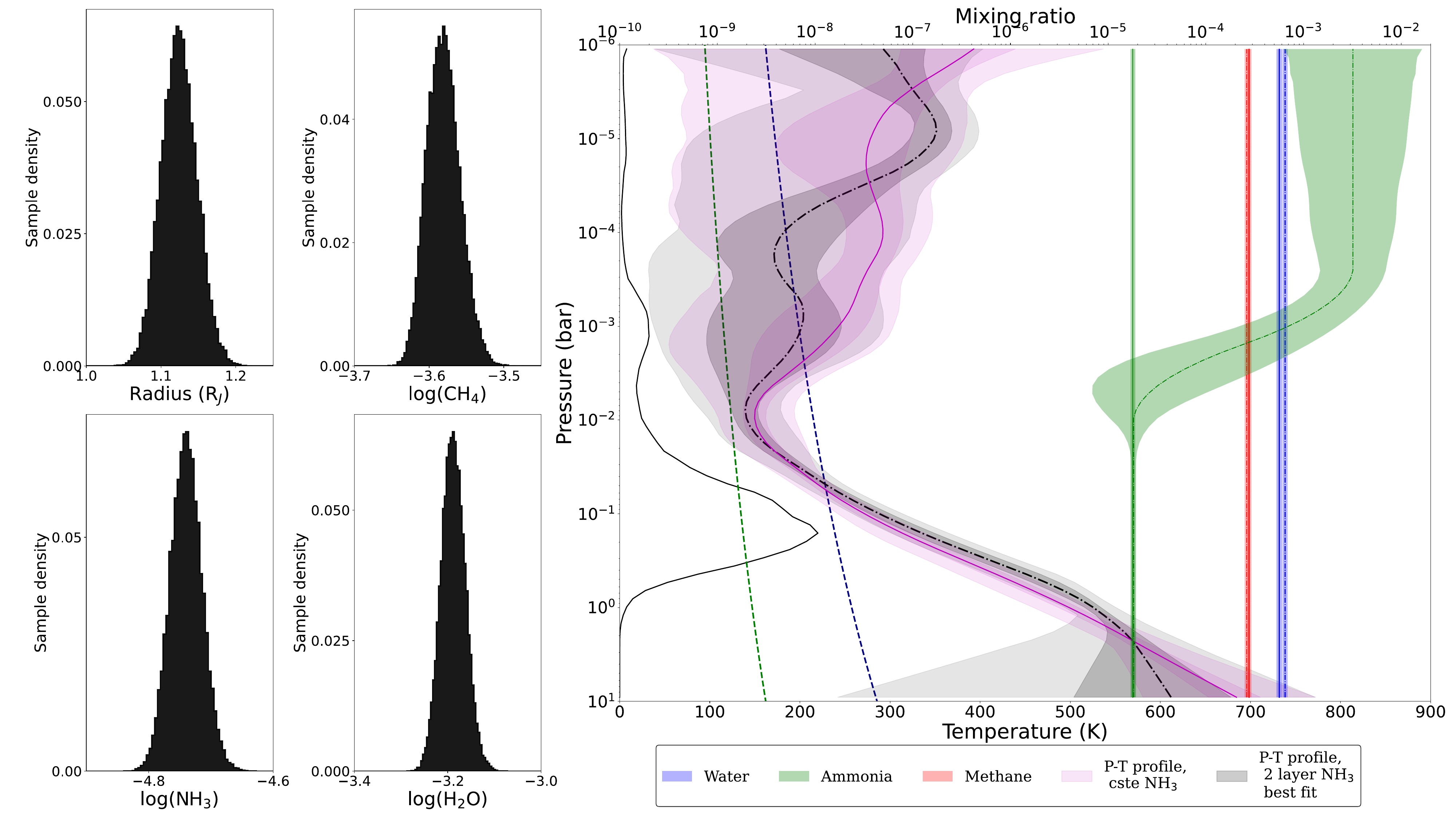}
\caption{Summary of the main retrieval results from our Best setup. Left panel: histograms of the final samples for the radius, mixing ratios of CH$_4$, H$_2$O and NH$_3$ from the best setup. 
Right panel: pressure temperature profile from the best setup (black) and for the baseline retrieval with constant NH$_3$ (purple), both corner plots are in Appendix \ref{Appendix:figures} Fig. \ref{fig:corner_best_setup} and Fig. \ref{fig:corner_cste_setup}}. The chemical profiles, with $1\sigma$ region, of Water (blue), Ammonia (green) and methane (red) with dashed line are from the best setup and the full lines from the baseline setup. The navy and green dashed line shows water and ammonia condensation curve for solar metallicity from \cite{2002Icar..155..393L}. The black full line is the radiative contribution function for the best setup.  \label{fig:profiles}
\end{figure*}

The pressure region $p = [1, 0.01]$\,bar comprises $\sim 81.6$ percent of our information content (see Fig. \ref{fig:profiles}). This region is therefore well constrained. Our retrievals yield a decreasing with altitude temperature profile, from $T = 547^{+39}_{-49}$\,K at $p = 1$\,bar to $T = 118^{+67}_{-57}$\,K at $p = 10^{-2}$\,bar. The probability density distributions for water, methane, and ammonia in Figure \ref{fig:profiles} exhibit mixing ratio values of $\rm{log(H_2O)} = -3.19^{+0.08}_{-0.07}$,   $\rm{log(CH_4)} = -3.58^{+0.07}_{-0.06}$ and $\rm{log(NH_3)} = -4.74^{+0.08}_{-0.07}$ with their $\pm 3 \sigma$ limits. These abundances all fall into the range observed for other Y$0$ dwarfs \citep{2019ApJ...877...24Z, Barrado_15NH3, 2024ApJ...971..121K}.

The contribution function shows that $18.1\%$ of the data informs us about the atmosphere at $p < 0.01$\,bar. The information at such pressures comes from the $\rm{NH_3, H_2O\,and\,CH_4}$ lines between $\lambda \in [5.7, 8.0]\,\mu$m and the two ammonia lines between $\lambda \in [10, 11]\,\mu$m. For $p \in [0.01, 10^{-4}]$\,bar, the $T-p$ profile is consistent with isothermal at the $1 \sigma$ (see black region on Fig. \ref{fig:profiles}).

At high altitude ($p < 10^{-4}$\,bar) and up to the top of our domain ($p = 10^{-6}$\,bar) the $T-p$ shows a negative lapse rate (i.e, thermal inversion). Specifically, the retrieved median model shows a $\sim 200\,$K increase from $ T = 154^{+89}_{-87}$\,K at $p = 10^{-4}$\,bar to $T = 381^{+37}_{-43}$\,K at $p = 10^{-5}$\,bar. This increase produces additional flux for the lines mentioned above. However, we note that this thermal inversion is weak, and the 3$\sigma$ temperature profile remains consistent with isothermal. Such thermal inversions could be present, even in a low irradiation regime such as WD\,0806\,b, due to e.g., breaking of gravity waves, or diabatic convection \citep{2019ApJ...882..117L, 2019ApJ...876..144T, 2023ApJ...959...86L}. \\

When using a two-layer profile for ammonia (see e.g., best setup, cloudy setup, 5 species setup and free setup) the high-altitude layer is systematically retrieved with a mixing ratio\footnote{proportion of the molecule in the gas} of $\rm{log(NH_{3~top})} = -2.5 ^{+1.5}_{-0.9}$ (see dashed green line on Fig. \ref{fig:profiles}). The mass retrieved is $\rm{M_p} = 0.51^{+0.57}_{-0.2}~\rm{M_J}$ ($3\sigma$ uncertainties). For the radius, our best setup retrieves $\rm{R_p} = 1.12^{+0.07}_{-0.07}~\rm{R_J}$, which is stable across the different setups mentioned above. We note that the planetary mass is outside the expected mass range from formation and evolutionary models. The ammonia gradient is also physically unexpected given the  $T-p$ range. These points are discussed more  in sections \ref{sec:Discussion:2} and \ref{sec:Discussion:4}.  Despite the retrieved $T-p$ profile crossing the condensation curve of water (see Fig. \ref{fig:profiles}), our retrieval does not find evidence for clouds, providing us with an upper limit of $\rm{log(P_{clouds})} > 3.2^{+6.8}_{-2.4}$ bar.\\

Except H$_2$O, NH$_3$, and CH$_4$, we do not find evidence for other major opacity sources. We do not detect $\rm{CO}$ and $\rm{CO_2}$: the 5-species setup infers upper limits of log(CO) $< -5.9$ and log(CO$_2$)$ <-5.3$. These molecules do not have significant absorption lines in the range of MIRI-LRS, making their detection challenging.  However, $\rm{CO_2}$ has a strong feature at $15$ microns, so the photometric data can be used to obtain a stricter constraint. We search for such a stricter constraint by increasing the mixing ratio of $\rm{CO_2}$ in our best-fit model until its absorption leads to $5\sigma$ departure in the $\rm{F1500W}$ filter. That method provides a mixing ratio upper limit of log(CO$_2$) $ < -6.8$. 

In the four hours spanning our LRS observations, we do not detect significant planetary variability: all our integrations are compatible with each other (see Fig. \ref{fig:variability}). Additionally, we note that the photometric data were taken about one day before the MIRI-LRS data, and they appear consistent with our best-fit forward model. Hence, the emission from the planet's atmosphere appears stable over the 24-hour period of our observations.

\section{Discussion} \label{sec:Discussion}

\subsection{Clouds}

Extending our best-fit model to $\lambda \in [0.3, 51.4]~\mu$m\footnote{The molecule's cross-sections wavelength ranges}, we compute the effective temperature of WD\,0806\,b to be $T_\mathrm{eff} = 343\pm11$\,K. Hence, we explored the possibility of condensation for the species we detect, i.e. $\rm{H_2O}$,  $\rm{NH_3}$, and $\rm{CH_4}$. Using the condensation curve from \cite{2002Icar..155..393L} and our retrieved P-T profile, we determine that only water is likely to condense while ammonia's condensation is less probable (see Fig. \ref{fig:profiles}). The condensation curve from \cite{2002Icar..155..393L} depends on the atmosphere's metallicity. As we do not have access to that parameter, we use a range $Z \in [0.1, 10]\cdot Z_\odot $, to estimate the pressure range where water should condense. We derived that water should condense at a pressure between $p \in [3.5\cdot10^{-2}, 6.2\cdot10^{-2}]$ depending on the metallicity. However, $60 \%$ of our information comes from below $6\cdot10^{-2}$ bar. Hence, if water condenses and forms significant clouds, we should detect it. 

We searched for water and ammonia ice cloud signatures by adding their absorption properties using the Mie scattering code \textsc{PyMieScatt} \citep{SUMLIN2018127}. This is included in \textsc{TauREx} via the \textsc{TauREx-PyMieScatt} plugin (\ref{sec:methodo:3}). This corresponding cloudy setup reached $\rm{ln(E)} = 10386$, a positive difference of $1$ compared to our best setup, which is not significant \citep{bayes_factor}. We therefore prefer the simpler cloudless model with 8 less free parameters \citep{2008ConPh..49...71T}. To further push our investigations, and considering that clouds might not extend in a sufficiently wide pressure range to produce major spectral features on our emission spectrum \citep{2007JOSAB..24..126H, article},  we also study a rainfall mechanism that sequesters water from the gas phase. This is implemented by using a two-layer profile for water (free water setup). The retrieved mixing ratio profile for water in this case is unconstrained at $p < 0.01$\,bar. The retrieved profile is consistent with a constant with altitude mixing ratio and  $\Delta \rm{ln(E) = -1.3}$, hence, we prefer the simpler model with a constant water profile \citep{2008ConPh..49...71T}. Note that we attempted to force a high-altitude water depletion by fixing log(H$_2$O$_\mathrm{top}) = -12$ and obtained $\Delta \rm{ln(E) = -1.7}$. Again, a constant mixing ratio for water is preferred. 

Our retrievals don't provide significant evidence for the presence of clouds in the atmosphere of WD\,0806\,b. Our observation and retrievals could be compatible with a rain fall mechanism involving a thin cloud layer. However, as with the clouds, a simpler constant with altitude abundance is statistically favored. While MIRI-MRS could probe lower pressure range, this source is too faint \citep{2024arXiv241010933K, Barrado_15NH3}. NIRSpec could access the $\lambda = 3.5\,\mu \rm{m}$ water ice feature and help us understand the formation of clouds in exoplanets.

\subsection{Robustness of the $\rm{NH}_3$ gradient and relation to the mass estimates}\label{sec:Discussion:2} 

 Below $0.01$\,bar, all retrievals systematically converge towards a mixing ratio consistent with $\rm{log(NH}_3^{surf}) = -4.74^{+0.08}_{-0.07}$\footnote{This is the value from the best setup with $3\sigma$ errors.} as shown in Figure \ref{fig:corner_best_setup}. However, we achieve the highest Bayesian evidence for retrievals with a positive ammonia gradient through the atmosphere. This gradient is statistically significant as the difference in Bayesian evidence between the best setup and the baseline setup is $\Delta \rm{ln(E)}=23.5$. The best setup favors a mixing ratio at the top of the atmosphere at $\rm{log(NH}_3^{top}) = -2.5 ^{+1.5}_{-0.9} $. However, we did not find a physically convincing explanation for this increase in the ammonia mixing ratio. The retrieved $T-p$ profile does not allow for the $\rm{N}_2$/$\rm{NH}_3$ conversion reaction to be significant, and the flux from the white dwarf is estimated to be too weak ($\approx 10^{-9}$ W) to significantly induce photochemical reactions. \\

 Besides the ammonia chemistry, the major change arising from having a two-layer ammonia profile is a stricter constraint of the planet's mass. For example, the best setup yields $\rm{M_P} = 0.51^{+0.57}_{-0.2}$ whereas the baseline setup gives $\rm{M_P} = 0.75^{+1.0}_{-0.30}~\rm{M_J}$ ($\pm 3\sigma$). This can easily be explained because the mass changes the pressure scale of the planet. The pressure of the ammonia gradient adds a constraint on the pressure scale compared to a constant mixing ratio of ammonia. Hence, as the pressure scale is better constrained, the mass constraint is stricter. Using retrievals, we show that higher masses result in the gradient located at higher pressures, whereas lower masses result in a $\rm{NH_3}$ gradient higher in the atmosphere. At the $3 \sigma$ level, the gradient's mid-pressure is constrained between log(p$_\mathrm{NH_3}$)$ \,= 2.2^{+0.7}_{-3.1}~\rm{Pa}$.

 As this ammonia gradient is statistically significant we report it but, given the lack of physical explanation,  this point should be cautioned.
 Given the correlation between the mass constraint and the ammonia gradient, we prefer to use the wider mass estimates from the baseline setup. 

\subsection{Element ratios}

WD\,0806\,b orbits a white dwarf on a wide separation (i.e., semi major axis,  $\rm{a}  \geq 1300$ au), unknown orbit. We derive the atmospheric elemental ratios to compare WD\,0806\,b with other known Y$0$ dwarfs.  As the host star is a white dwarf, we do not have access to the progenitor main-sequence star's element ratios  \citep{Subasavage_wd0806_age}. Hence, we cannot easily use the following ratio to conclude on the probable formation path of the planet. Nevertheless, from the water, ammonia and methane abundances reported in figure \ref{fig:profiles} we derive estimates for C/O, C/N and N/O. Using \cite{2002Icar..155..393L} and given our retrieved $T-p$ profile, we establish that $\rm{NH_3}$, $\rm{H_2O}$ and $\rm{CH_4}$ are the  major gas bearing species for N, O and C respectively. \cite{2023MNRAS.520.4683F} showed that oxygen entrapment in refractory species changes with equilibrium temperature and metallicity. However, as our temperature range is not covered, we settle on a middle value of $20 \%$ entrapment in refractory species between the estimate of \cite{2002Icar..155..393L} and \cite{2023MNRAS.520.4683F}. For the nitrogen content, we do not consider the increase in ammonia at high altitude given the low pressure, the amount of added nitrogen is negligible.  We derive $\rm{C/O} = 0.34 \pm 0.06$, $\rm{C/N} = 14.4 ^{+2.5}_{-1.8}$ and $\rm{N/O} = 0.023 \pm 0.004$. The derived $\rm{C/O}$ ratio is similar to those observed in known Y0 dwarfs \citep{2019ApJ...877...24Z,Barrado_15NH3,2024ApJ...971..121K, 2024ApJ...976...82T}.Moreover, using $\rm{Z} = \rm{(C + N + O)}/ \rm{H}$ as a proxy for metallicity, we find a slightly sub-solar metallicity. More specifically,  $\rm{Z} = 0.74^{+0.13}_{-0.09} \cdot \rm{Z^{\rm{CNO}}_{\odot}}$ with $\rm{Z^{\rm{CNO}}_{\odot}} =  \rm{(C_{\odot} + N_{\odot} + O_{\odot})}/ \rm{H_{\odot}}$ \citep{2009ARA&A..47..481A}.

\subsection{Consistency of the mass and radius with evolutionary models}\label{sec:Discussion:4}

We use the models from \cite{2021ApJ...920...85M} to obtain another estimate of WD\,0806\,b's mass from evolutionary models. Using the retrieval independent system's age quoted in \cite{Luhman_wd-0806_confirm}, $2\pm0.5$\,Gyr, and the effective temperature of $T_\mathrm{eff} = 343\pm11$\,K computed by integrating the flux of our forward model, we obtain a predicted planetary mass of $\rm{M_p} \in [6.3, 9.4]\,\rm{M_J}$. This is inconsistent with our retrieved mass of $\rm{M_p} \in [0.45, 1.75]~\rm{M_J}$ (see table \ref{tab:priors}). Note that our estimated effective temperature is one of the most robust parameters we derive and matches the findings from \cite{Luhman_wd-0806_confirm} obtained via Spitzer photometry. {We identify three possible sources for the above discrepancies in radius, mass, and effective temperature. \\

1) Our retrieved mass is incorrect. Assumptions made in the retrievals -- such as the opacity source treatment, 1D plane-parallel atmosphere, or free chemistry -- could bias the retrieved value of the mass. \\

2) The age used for evolutionary model predictions is incorrect. If the retrieved mass is correct, it matches the evolutionary track of a much younger object ($60 - 180$ Myr old). Such a young age would require that WB\,0806\,b formed after its host star, as could be the case of a captured object. \\

3) Assumptions of the evolutionary models do not apply to this case. The evolutionary models could be neglecting important opacity sources, affecting cooling rates \citep{2007ApJ...661..502B}. Other possibilities could include unaccounted energy sources, or the planet being a binary \citep{2024arXiv241011953X}. We are able to exclude tidal heating as a likely additional heating source, as the periapsis is at least $10$ au\footnote{We used the most favourable orbital parameters with an eccentricity and a of $0.99$ and $1300$ au respectively.}. Additionally, the stellar flux is too weak ($\approx 10^{-9}$ W) for the stellar irradiation to be a sufficient source of heat. \\

The retrieved radius ($\rm{R_p} = 1.12^{+0.07}_{-0.07}~\rm{R_J}$, see Table \ref{tab:priors}) is consistent with a planet of  $2\pm0.5$ Gyr old and a mass of $\rm{M_p} \in [6.3, 9.4]\,\rm{M_J}$ (i.e., the retrieved mass would be incorrect), but also with a much younger ($60 - 180$\,Myr) and lighter planet with $\rm{M_p} \in [0.45, 1.75]~\rm{M_J}$ (i.e., the age estimate would be incorrect). 

A stricter constraint on the radius, $1\% $ instead of the current $6.5\%$, could help distinguishing the young object scenario from the old one. The joint retrieval analysis of both MIRI and NIRSpec could help provide such stricter constraint.

\section{Conclusion} \label{sec:cite}

We reduced the MIRI-LRS data of WD\,0806-661\,b, modifying the background subtraction step, adding a new background rescaling step and using a custom-made extraction. The \textsc{TauREx} fit of the extracted spectrum yields precise constraints on the water, methane, and ammonia mixing ratio, as well as the $T-p$ profile of this atmosphere. While the $T-p$ profile allows for condensation of water, our search for signatures of aerosols did not favor this scenario: a cloud-free model is preferred. Using the forward model, we compute the precise effective temperatures of the planet and place upper limits on the abundance of carbon monoxide and carbon dioxide. Our retrievals yield a smaller than expected planetary mass, which is at odds with evolutionary models tracks based on the white dwarfs' progenitor's age. However, this smaller mass is consistent with a younger object formed later than the host's system. The radius we derive is consistent with both scenarios, following analysis with additional NIRSpec data will help us select the correct explanation. Further study, potentially using more self-consistent atmospheric models, or relying on complementary data with NIRSpec can allow us to assess the robustness of these parameters and place further constraints on $\rm{CO}$ and $\rm{CO_2}$ abundances.

\section{Acknowledgments}
  
The authors thank the anonymous referee for helpful comment, improving the quality of the paper. The authors thank G. Sloan, B. Trahin, M. Regan and the other STScI experts for their precious inputs and knowledge of MIRI and the STScI pipeline. M. Voyer extends his gratitude to L. Pueyo for providing funds for a two-month collaboration at STScI and S. Schleich as well as A. Dyrek for insightful discussions on retrieval methods and pipeline development. M. Voyer is also grateful to E. Matthews for her help to find the Sonora models. P-O. Lagage and M. Voyer acknowledge funding support by CNES. This publication is part of the project "Interpreting exoplanet atmospheres with JWST" with file number 2024.034 of the research programme "Rekentijd nationale computersystemen" that is (partly) funded by the Netherlands Organisation for Scientific Research (NWO) under grant https://doi.org/10.61686/QXVQT85756. This work used the Dutch national e-infrastructure with the support of the SURF Cooperative using grant no. 2024.034. D. Barrado is supported by Spanish MCIN/AEI/10.13039/501100011033 grants PID2019-107061GB-C61 and PID2023-150468NB-I00. O. Absil is a Senior Research Associate [Research Director] of the Fonds de la Recherche Scientifique – FNRS. O. Absil thanks the European Space Agency (ESA) and the Belgian Federal Science Policy Office (BELSPO) for their support in the framework of the PRODEX Programme. J. Pye acknowledges financial support from the UK Science and Technology Facilities Council, and the UK Space Agency. G. \"Ostlin acknowledges support from the Swedish National Space Agency (SNSA). This work is based on observations made with the NASA/ESA/CSA James Webb Space Telescope. The data were obtained from the Mikulski Archive for Space Telescopes at the Space Telescope Science Institute, which is operated by the Association of Universities for Research in Astronomy, Inc., under NASA contract NAS5-03127 for JWST. These observations are associated with the program $\#1276$. The specific observations analyzed can be accessed via \dataset[doi: 10.17909/anme-j453]{https://doi.org/10.17909/anme-j453}. This project was provided with computing HPC and storage resources by GENCI at TGCC thanks to the grant 2024-15722 on the supercomputer Joliot Curie’s SKL and ROME partition. For the purpose of open access, the authors have applied a Creative Commons Attribution (CC BY) license to the Author Accepted Manuscript version arising from this submission.

\vspace{5mm}
\facilities{JWST (MIRI LRS Fixed Slit and MIRI Imager)}

\software{TauREx \citep{Al-Refaie_taurex3_2021, Al-Refaie_taurex3_2022}, MultiNest \citep{Feroz_MultiNest_2009, Buchner_MultiNest_2014}, PyMultiNest \citep{Buchner_MultiNest_2014}, JWST Pipeline \citep{Bushouse_JWST_Calibration_Pipeline_2024}, Photutils  
\citep{larry_bradley_2024_12585239}.}

\bibliography{main}{}
\bibliographystyle{aasjournal}

\vfill
\clearpage

\appendix

\section{Appendix 1: Pipeline modifications}\label{Appendix:A}
\subsection{Background subtraction}\label{Appendix:A1}

The pipeline -- version 1.12.5 -- we use does not include error propagation from the background subtraction step\footnote{\url{https://jwst-pipeline.readthedocs.io/en/latest/jwst/error_propagation/main.html}}. We therefore create a custom background subtraction step to include this additional error term. We start by building the background images with the half images, from the “rateints” files,  that does have the target signal (like the right-hand side of Fig. \ref{fig:appendixA2} left panel). Scanning pixel by pixel, to perform a $5 \sigma$ clip on the integrations to reject temporal outliers. If a value is flagged, it is replaced by the mean of the integrations. If an NaN (white pixels in Fig. \ref{fig:appendixA2} left panel) \footnote{We keep the warm and hot pixels as they are present in the other exposure.} is present in all integrations, we interpolate it. To that effect, we average the integrations together to fit a second order \footnote{second order needed to fit the shape of the background gradient.} polynomial to the line (cross dispersion direction) and recover the value. The error of these recovered pixels is estimated as the average of the two neighboring pixels' error (in the cross dispersion direction). We use this time-outlier and NaN free averaged-image from one exposure for our background subtraction. This background image from one exposure is subtracted to the integrations of the other exposure, with proper error propagation.

\subsection{Outlier interpolation of the science images}\label{Appendix:A2}

\begin{figure*}[ht!]
\includegraphics[width=1\textwidth]{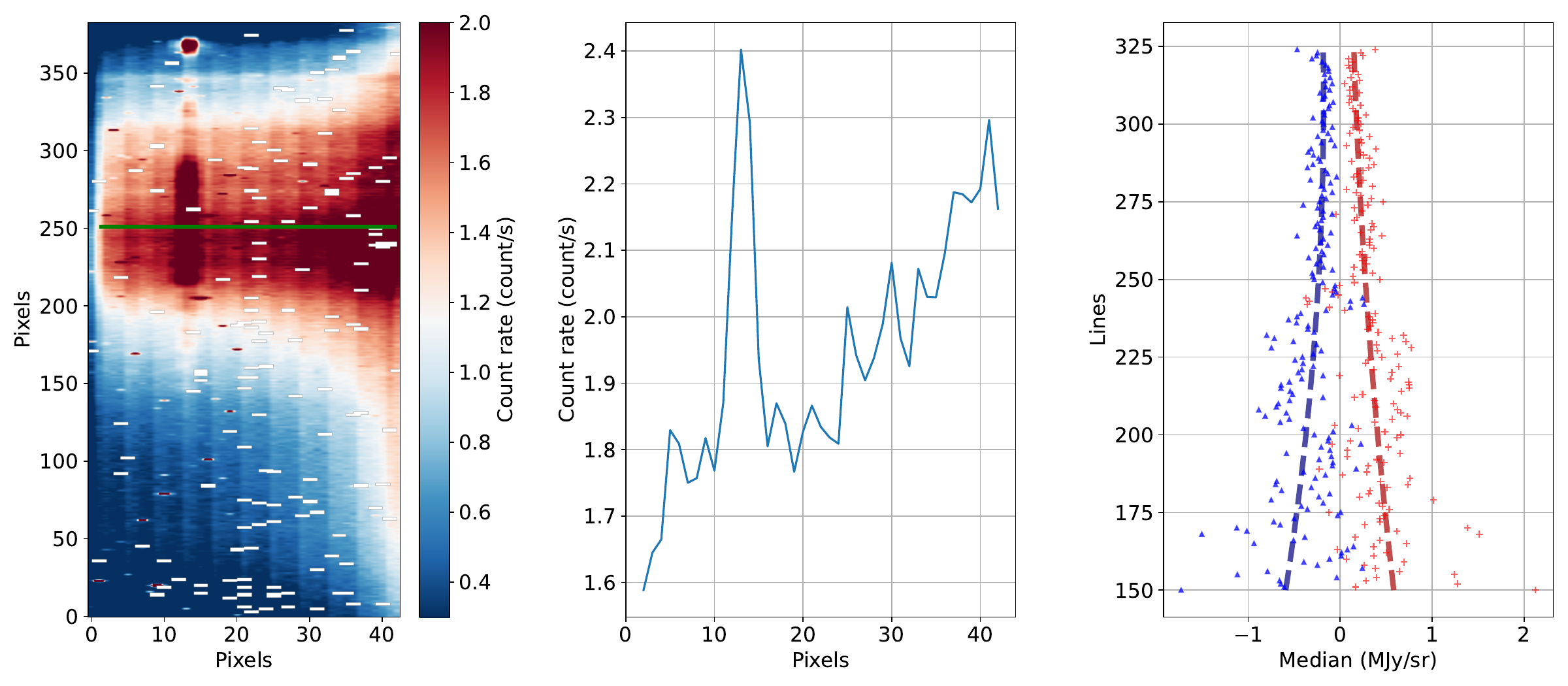}
\caption{Left panel: Mean image of the first exposure after the ramp fitting. Middle panel: plot of the green line shown in the left panel. We clearly see the right to left gradient, and the target's signal. Right panel: line by line median for each of the two exposures and associated second order polynomial fit.  \label{fig:appendixA2}}
\end{figure*}

After stage $2$ we remove the temporal and spatial outliers from the science images. This step begins by identifying temporal outliers within each exposure (eight integrations per exposure, so we compare eight values). For each value, we calculate the median and standard deviation (std) by excluding one value at a time from the eight-value set, performing this calculation eight times. We then compute the median of these medians and standard deviations, which we refer to as the “master median” and “master std.” Using these metrics, we apply a $5\sigma$ clip to the full set of eight values, removing any outliers. This method offers improved robustness over a standard $5\sigma$ clip, which can be skewed by extreme outliers. The identified outliers are replaced with the mean of the remaining dataset.

Our data contains a number of spatial outliers (i.e., outliers present on every integration, isolated dark red points in Fig. \ref{fig:appendixA2}). They are detected using a $5 \sigma$ clip with their eight direct neighbors ($3\rm{x}3$ corona). Such outliers can create charge spillage effect, where they affect the values of their direct neighbors. We perform for each of the center pixel direct neighbors a  $5 \sigma$ clip on the $5\rm{x}5$ corona to detect such contamination and replace all fagged pixels with NaNs.  

\subsection{Background rescaling}\label{Appendix:A3}

As seen in figure \ref{fig:flux_offset} the flux levels of the source in the different exposures do not match. To correct this, we introduce an additional “Background rescaling” step. Working on the “calints” files after the outlier interpolation exposed above in Appendix \ref{Appendix:A2}, we mask both positive and negative traces (image area impacted by the target's signal), with an 11 pixel wide rectangle. Given that the images are 42 pixels wide and that we use two, $11$ pixels wide mask per image, each remaining line has 20 pixels. To perform a robust statistical treatment, we merge the arrays of the 8 integrations together so that the combined lines are $8 \cdot 20 = 160 $ pixels wide. For each line of these $160$ pixel wide images, we compute the line-by-line medians in the cross-dispersion direction (e.g., see right panel of Fig. \ref{fig:appendixA2}). Those medians, therefore, capture the wavelength dependent variation of the background's gradient in the two exposures (i.e., corresponding to the two source positions). A second order polynomial is then fitted to these medians (right panel of Fig. \ref{fig:appendixA2}) to smooth our background solution. Finally, line by line, the associated polynomial's value is subtracted to the pixels' signal.

\subsection{Spectral extraction}\label{Appendix:A4}

To extract the spectra from the individual integrations, we utilize the “calints” files, after correction by our “outlier interpolation” and our “Background Rescaling” step described in Appendix \ref{Appendix:A2} and \ref{Appendix:A3}. We utilize a six-pixel-wide rectangular aperture or a four-pixel-wide aperture if NaNs are present in the two edge pixels. When NaNs occur within the central four pixels, we interpolate them using a Gaussian function (on the cross-dispersion direction). For this Gaussian interpolation -- because of the curved trace -- we fix the mean of the Gaussian function to the center of the trace, which is obtained by fitting a second order polynomials to the NaN-free lines. If the Gaussian has a standard deviation smaller than the length of the line, we use it to interpolate the missing value. However, if multiple NaNs are present in the center of the trace, we discard the entire line, as there is insufficient information to reliably recover the missing values. In that case, the data point on the final spectrum will come from only one exposure (the spectrum error also reflects this).
Lastly, we extract accounting for the finite aperture using the standard “$\rm{jwst\_miri\_apcorr\_0012}$” file. The integrations are then combined, followed by the exposures, to produce the final spectra presented in figure \ref{fig:LRS&fit}.

\newpage

\section{Appendix: Wavelength solution}\label{Appendix : wavelength correction}

This section contains the table with the adopted wavelength correction for this study.

\begin{table}[ht]
\centering
\begin{tabular}{c|c|c}

\multicolumn{2}{c|}{Row number} & \multirow{2}{*}{Wavelength ($\mu m$)}\\

MIRIm coordinates & LRS coordinates & \\ \hline \hline

393 & 1 & 3.12478 \\
392 & 2 & 3.62459 \\
391 & 3 & 3.79141 \\
... & ... & ... \\
8   & 386 & 13.95934 \\
7   & 387 & 13.97298 \\
6   & 388 & 13.98651 \\

\end{tabular}
\caption{Row number with the associated wavelength adopted in this study. Row numbers, labeled $\rm{Yrow}$, are given in both MIRIm coordinates ($\rm{Yrow} \in [1, 2024]$) and  LRS Slit coordinates from calibration file "jwst\_miri\_specwcs\_0008.fits ($\rm{Yrow} \in [1, 388]$). The full table is available in the journal version of this paper.}
\label{tab:wavelength_correction}
\end{table}

\section{Appendix: retrieval priors}\label{Appendix : B}

Here we present Table \ref{tab:priors}, which contains the priors and retrieved parameters for the best setup and baseline setup retrievals.

\begin{table}[ht!]
    \centering
    \begin{tabular}{c|c|c|c}  
         \multirow{2}{*}{Parameters} &\multirow{2}{*}{Priors} & \multicolumn{2}{c}{Retrieved parameters}\\
 & & Best setup&Baseline setup\\ \hline  \hline 
         $\rm{R_P}$&  $\mathcal{U}\in[0.5 - 2]\, \rm{R_P}$&$1.12^{+0.07}_{-0.07}~\rm{R_J}$ &$1.18^{+0.08}_{-0.08}~\rm{R_J}$\\   
         $\rm{M_P}$&  $\mathcal{U}\in[0.3 - 10]\,\rm{M_P}$&$0.51^{+0.57}_{-0.2}~\rm{M_J}$ &$0.75^{+1.1}_{-0.3}~\rm{M_J}$\\   
         $\rm{T_{Surf}}$&  $\mathcal{U}\in[20 - 1000]\,\rm{K}$&$615^{+169}_{-389} \rm{K}$ &$693^{+94}_{-110} \rm{K}$\\   
         $\rm{T_1}$&  $\mathcal{U}\in[20
- 1000]\,\rm{K}$&$547^{+39}_{-49} \rm{K}$ &$504^{+30}_{-47} \rm{K}$\\   
         $\rm{T_2}$&  $\mathcal{U}\in[20
- 1000]\,\rm{K}$&$227^{+30}_{-35} \rm{K}$ &$261^{+24}_{-39} \rm{K}$\\   
         $\rm{T_3}$&  $\mathcal{U}\in[20
- 1000]\,\rm{K}$&$118^{+67}_{-57}\,\rm{K}$ &$126^{+75}_{-55}\,\rm{K}$\\   
         $\rm{T_4}$&  $\mathcal{U}\in[20
- 1000]\,\rm{K}$&$215^{+127}_{-195} \rm{K}$ &$250^{+101}_{-229} \rm{K}$\\   
 $\rm{T_5}$& $\mathcal{U}\in[20
- 1000]\,\rm{K}$&$ 154^{+89}_{-87}\,\rm{K}$ &$ 305^{+101}_{-279}\,\rm{K}$\\
 $\rm{T_6}$& $\mathcal{U}\in[20
- 1000]\,\rm{K}$&$ 381^{+37}_{-43}\,\rm{K}$ &$ 263^{+170}_{-242}\,\rm{K}$\\
 $\rm{T_{Top}}$& $\mathcal{U}\in[20
- 1000]\,\rm{K}$&$289^{+51}_{-122}\,\rm{K}$ &$399^{+163}_{-373}\,\rm{K}$\\   
 $\rm{NH_{3\,Surface}}$& $\mathcal{U}\in[10^{-8}
- 10^{-1}]$&$-4.74^{+0.08}_{-0.07}$ &$-4.74^{+0.08}_{-0.06}$\\   
 $\rm{NH_{3\,Top}}$& $\mathcal{U}\in[10^{-8}
- 10^{-1}]$&$-2.5 ^{+1.5}_{-0.9}$ &Not in model\\   
 $\rm{NH_{3\,P}}$& $\mathcal{U}\in[10^{6}
- 0.1]\,\rm{Pa}$&$ 2.2^{+0.7}_{-3.1}~\rm{Pa}$ &Not in model\\   
 $\rm{CH_4}$& $\mathcal{U}\in[10^{-8}
- 10^{-1}]$&$ -3.58^{+0.07}_{-0.06}$ &$ -3.56^{+0.08}_{-0.06}$\\   
 $\rm{H_2O}$& $\mathcal{U}\in[10^{-8}
- 10^{-1}]$&$ -3.19^{+0.08}_{-0.07}$ &$ -3.25^{+0.1}_{-0.08}$\\   
 Flat Mie& $\mathcal{U}\in[10^{6}
- 0.1]\,\rm{Pa}$& $> 5.5^{+0.5}_{-0.6}\,\rm{Pa}$&$> 5.6^{+0.4}_{-0.5}\,\rm{Pa}$\\  
    \end{tabular}
    \caption{Priors used for our retrievals and best setup and baseline setup posteriors with $3\sigma$ limits. Molecules volume mixing ratios and pressures in $\rm{log_{10}}$ scale.} 
    \label{tab:priors}
\end{table}

\newpage

\section{Appendix: Additional figures}\label{Appendix:figures}
 This appendix contains the complementary figures to section \ref{sec:Results}, Figures \ref{fig:mol_depth}
 - \ref{fig:variability}.
\begin{figure*}[ht!]
\centering
\includegraphics[width=0.99\textwidth]{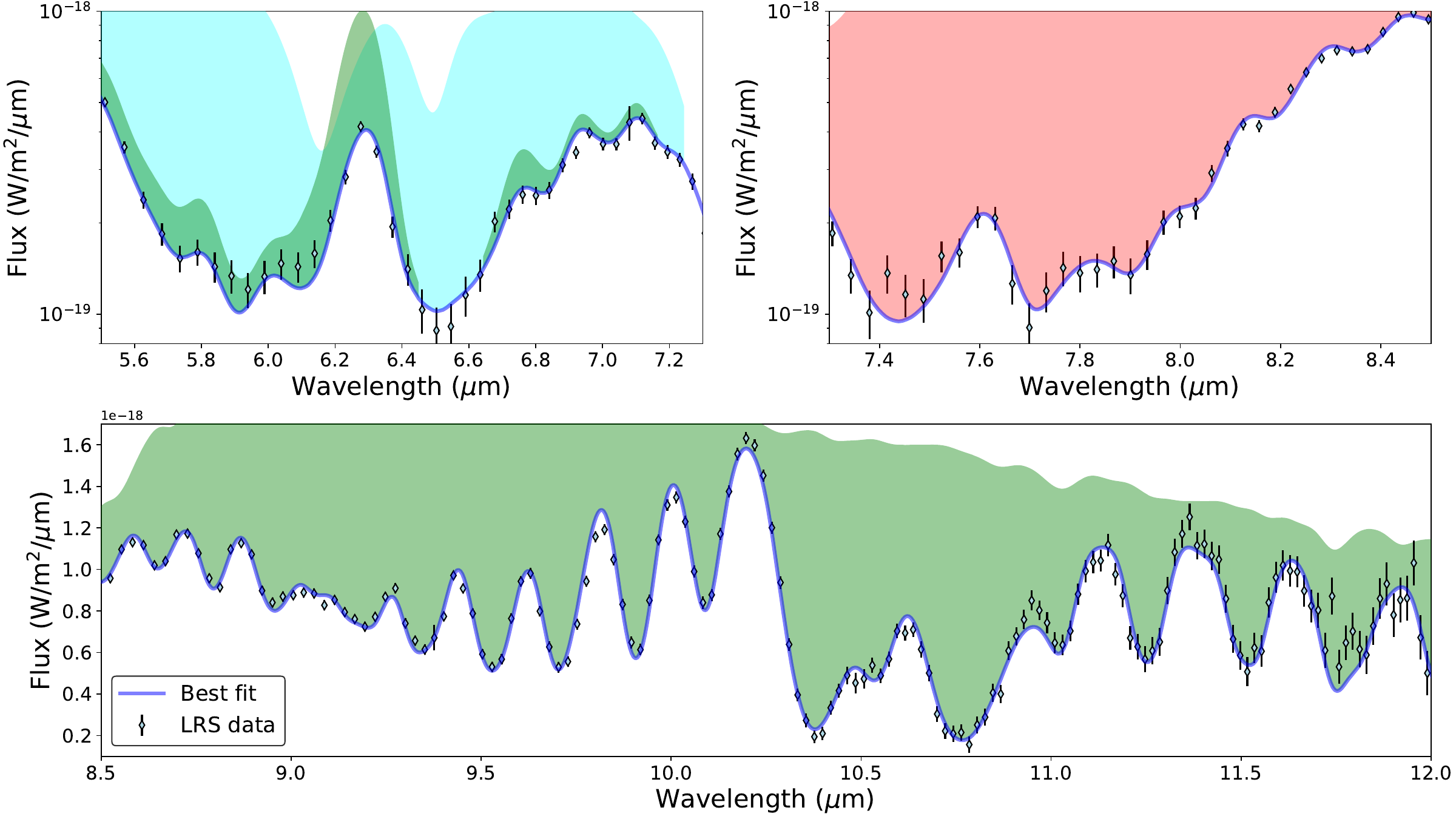}
\caption{Molecular absorption created by water (blue), ammonia (green) and methane (red). \label{fig:mol_depth}}
\end{figure*}
\newpage

\begin{figure*}[ht!]
\includegraphics[width=1\textwidth]{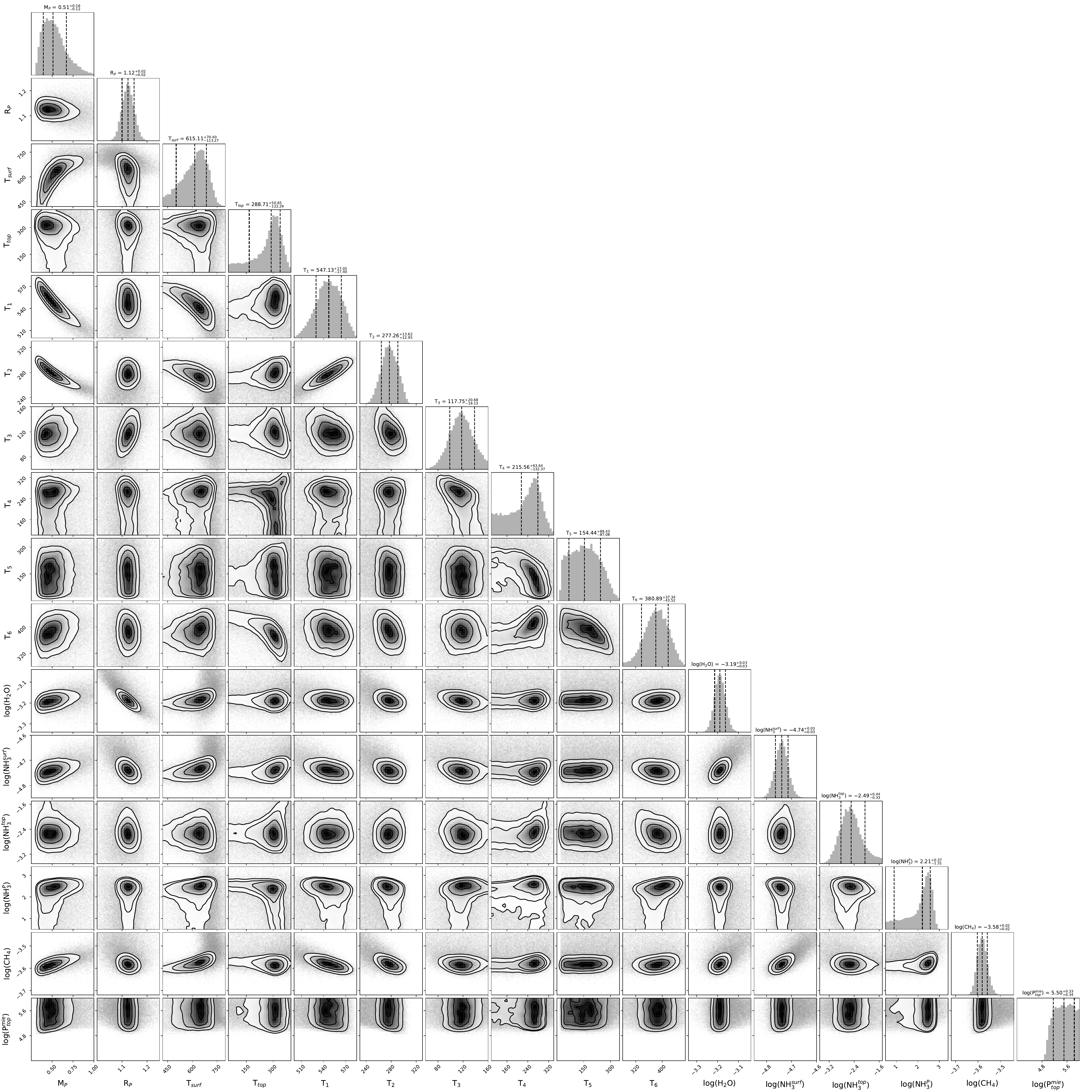}
\caption{Corner plot for our best setup. \label{fig:corner_best_setup}}
\end{figure*}
\newpage

\begin{figure*}[ht!]
\includegraphics[width=1\textwidth]{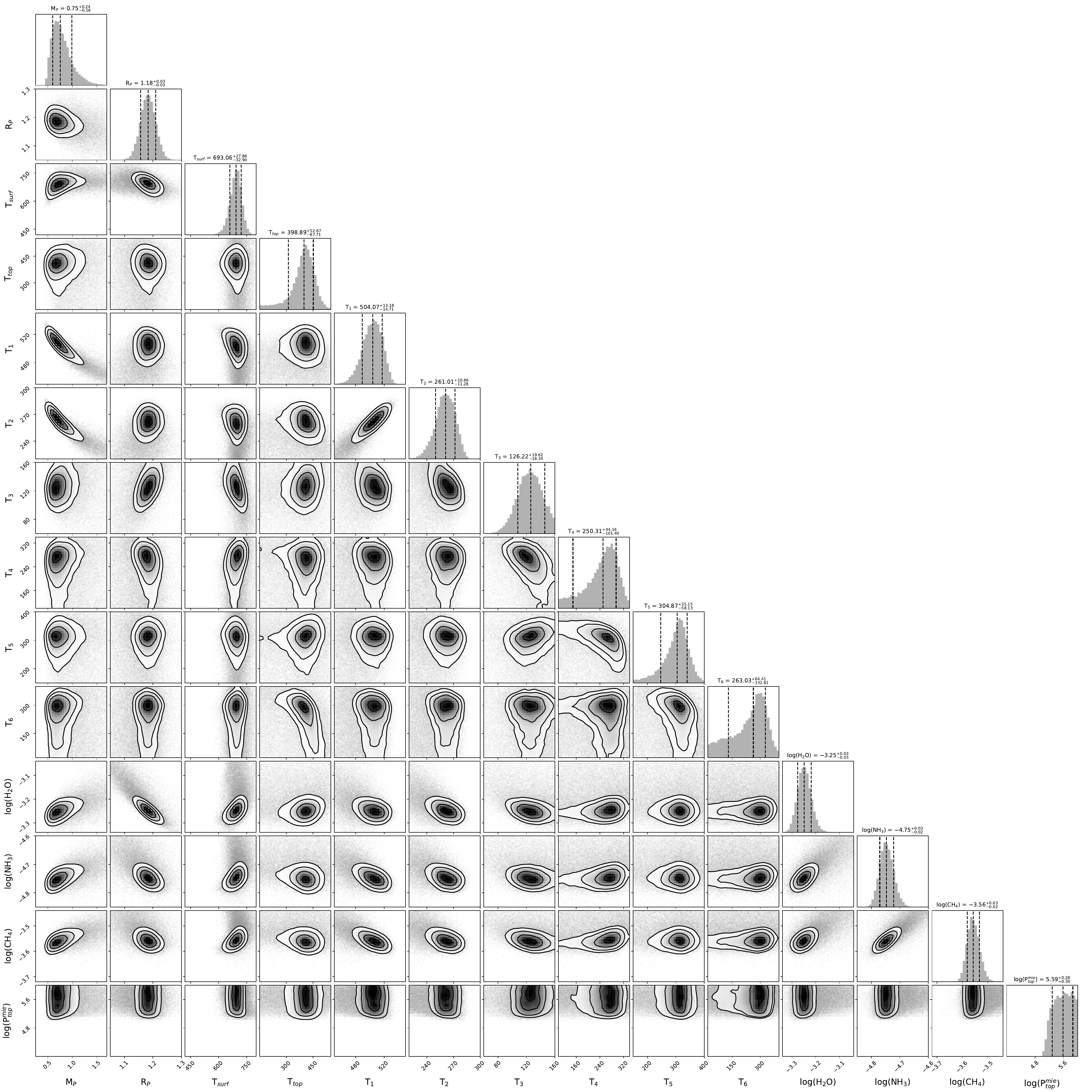}
\caption{Corner plot for our baseline setup. \label{fig:corner_cste_setup}}
\end{figure*}

\begin{figure*}[ht!]
\includegraphics[width=1\textwidth]{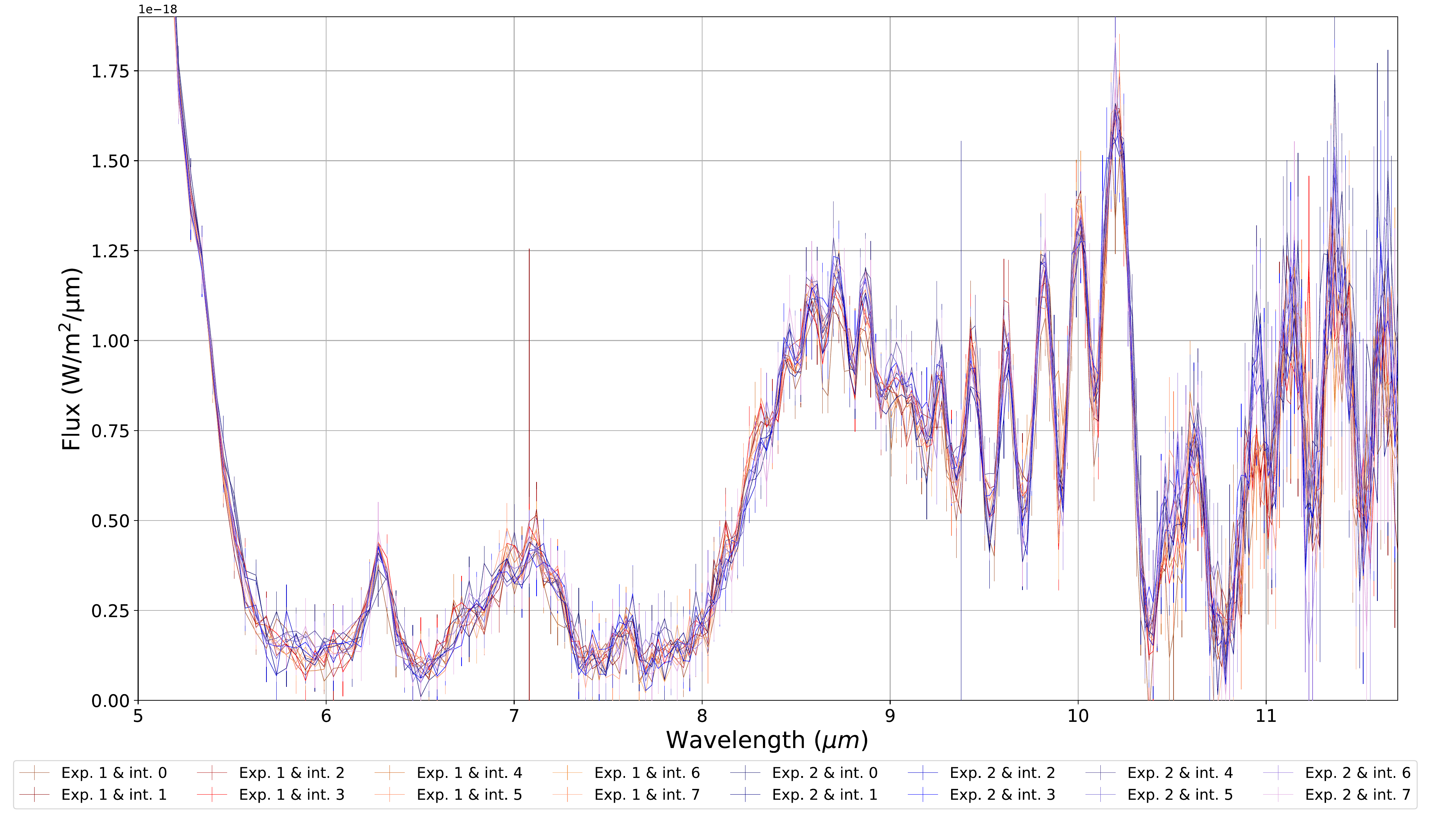}
\caption{Spectra extracted from every integrations after our “background rescaling” step. The compatibility of the integrations constrains the variability of the planet over the four-hour period they were taken.\label{fig:variability}}
\end{figure*}

\end{document}